\documentstyle[here,epsfig]{mn2e}
\begin{document}
\def\simlt{\mathrel{\rlap{\lower 3pt\hbox{$\sim$}}\raise 2.0pt\hbox{$<$}}}
\def\simgt{\mathrel{\rlap{\lower 3pt\hbox{$\sim$}} \raise
2.0pt\hbox{$>$}}}

\newcommand{\lsim}{\raisebox{-0.13cm}{~\shortstack{$<$ \\[-0.07cm] $\sim$}}~}
\newcommand{\gsim}{\raisebox{-0.13cm}{~\shortstack{$>$ \\[-0.07cm] $\sim$}}~}

\def\whs{\thinspace ${\rm W \: Hz^{-1} \: sr^{-1}}$}
\def\lpow{\thinspace ${\rm log_{10}P_{1.4GHz}\: }$}
\def\kv{\thinspace $K_{Vega}$}
\def\kab{\thinspace $K_{\rm AB}$}

\title[] 
{ The evolution of the near-IR galaxy Luminosity Function and colour bimodality up 
to ${\bf z \simeq 2}$ from the UKIDSS  Ultra Deep Survey  Early Data Release}

\author[M. Cirasuolo et al.] {
\parbox[h]{\textwidth}{M. Cirasuolo$^{1}$, R. J. McLure$^{1}$, J. S. Dunlop$^{1}$, 
O. Almaini$^{2}$, S. Foucaud$^{2}$, Ian Smail$^{3}$, 
K. Sekiguchi$^{4}$, C. Simpson$^{5}$,
S. Eales$^{6}$, S. Dye$^{6}$,
M.G. Watson$^{7}$, M.J. Page$^{8}$, P. Hirst$^{9,10}$ \\ } 
\vspace*{2pt} \\
$^{1}$SUPA\thanks{Scottish Universities Physics Alliance} Institute
for Astronomy, University of Edinburgh, 
Royal Observatory, Edinburgh EH9 3HJ\\
$^{2}$School of Physics and Astronomy, University of Nottingham,
University Park, Nottingham NG7 2RD\\ 
$^{3}$Institute for Computational Cosmology, Durham University, 
South Road, Durham DH1 3LE\\
$^{4}$Subaru Telescope, National Astronomical Observatory of Japan,
650 North A'ohoku Place, Hilo, Hawaii 96720, USA\\
$^{5}$Astrophysics Research Institute, Liverpool John Moores
University, Twelve Quays House, Egerton Wharf, Birkenhead CH41 1LD\\
$^{6}$ School of Physics and Astronomy, Cardiff University, 
5 The Parade, Cardiff, CF24 3YB \\
$^{7}$Department of Physics \& Astronomy, University of Leicester, 
Leicester LE1 7RH\\
$^{8}$Mullard Space Science Laboratory, University College London, 
Holmbury St. Mary, Dorking, Surrey RH5 6NT\\
$^{9}$Joint Astronomy Centre, 660 N. A'ohoku Place, University Park,
Hilo, Hawaii 96720, USA \\
$^{10}$ Gemini Observatory, 670 N. Aohoku Pl, Hilo, HI 96720, USA}
\maketitle 
\begin{abstract}

We present new results on the cosmological evolution 
of the near-infrared galaxy luminosity function, derived from the
analysis of a new sample of $\sim 22,000$ \kab $\le 22.5$
galaxies selected over an area of 0.6 square degrees 
from the Early Data Release of the UKIDSS Ultra Deep Survey (UDS).
Our study has exploited the multi-wavelength coverage of the UDS field 
provided by the new UKIDSS WFCAM $K$ and $J$-band imaging,
the Subaru/XMM-Newton Deep Survey and the {\it Spitzer}-SWIRE Survey. 
The unique combination of large area and depth provided by this new survey
minimises the complicating effect of cosmic variance and has allowed 
us, for the first time, to trace the evolution of the brightest 
sources out to $z \simeq 2$ with good statistical accuracy.

In agreement with previous studies we find that the characteristic 
luminosity of the near-infrared luminosity function brightens by 
$\simeq 1$ magnitude between $z=0$ and $z  \simeq 2$, while the total density 
decreases by a factor $\simeq 2$. Using the rest-frame $(U-B)$ colour to
split the sample into red and blue galaxies, we confirm the classic
luminosity-dependent colour bimodality at $ z \lsim 1$. However, the
strength of the colour bimodality is found to be a decreasing function
of redshift, and seems to disappear by $z \simgt 1.5$. Due to the
large size of our sample we are able to investigate the differing 
cosmological evolution of the red and blue galaxy populations.
It is found that the space density of the brightest red galaxies ($M_K
\le -23$) stays approximately constant with redshift, and that these
sources dominate the bright-end of the luminosity function at
redshifts $z\lsim 1$. In contrast, the brightening of the
characteristic luminosity and mild decrease in space density displayed
by the blue galaxy population leads them to dominate the bright-end of
the luminosity function at redshifts $z\gsim 1$.
\end{abstract}
%
\section{INTRODUCTION}

The potential of deep near-infrared observations to measure the cosmological
evolution of the stellar mass of galaxies has long been understood 
(e.g. Lilly \& Longair 1984; Dunlop et al. 1989; Glazebrook et al. 1995;
Cowie et al. 1996). Now, with the advent 
of wide-field near-infrared detectors, this potential has begun to be realized,
and several deep infrared-based studies supported by 
spectroscopic and/or photometric redshift information have recently been 
completed 
(e.g. Cimatti et al. 2002; Pozzetti et al. 2003; 
Dickinson et al. 2003; Drory et al. 2003;
Fontana et al. 2004; Glazebrook et al. 2004; Caputi et al. 2006; 
Drory et al. 2005). 

A key aim of this work is the determination of the cosmological 
evolution of the $K$-band galaxy luminosity function (LF), the present-day 
form of which is now reasonably well established (Cole et al. 2001; 
Kochanek et al. 2001). The most recent studies have been deep enough, 
and have been armed with sufficient redshift
information to attempt to trace this evolution out to $z = 2-3$. The results
indicate a substantial brightening in characteristic magnitude 
from $z =0 $ up to $z = 2-3$ along with a simultaneous decrease in galaxy 
number density (see e.g. Caputi et al. 2006; Saracco et al. 2006). 
A similar combination of luminosity and density evolution has also been 
observed in the optical LF (e.g. Lilly et al 1995; 
Poli et al. 2003; Gabasch et al. 2004; Giallongo et al. 2005). 

However, even these most recent deep near-infrared 
surveys have still only covered relatively small areas of sky 
(from a few $\rm arcmin^2$ in the Hubble Ultra 
Deep Field to several hundred $\rm arcmin^2$ 
in the Great Observatories Origins Deep Survey (GOODS))
and are hence existing results on the evolving LF are 
still potentially heavily affected by 
cosmic variance. In this paper we 
exploit the order-of-magnitude improvement in areal 
coverage (0.6 sq. deg.) provided by the 
the UKIRT Infrared Deep Sky Survey (UKIDSS; Lawrence et al. 2006) 
Ultra Deep Survey (UDS) to constrain the evolution of the near-infrared 
LF and the colour bimodality up to $z \simgt 2$.

The rest of this paper is structured as follows.
In Section 2 we summarize the main properties of the datasets 
used in this work. Then, in Section 3 we 
describe in detail how our new $K$-band galaxy sample 
has been constructed and cleaned, and describe the method 
we have developed for the estimation of redshifts from the 
available multi-band photometry. 
In Section 4 we assess and discuss the reliability of our 
photometric redshift estimates and in Section 5 we derive the near-infrared 
LF. 
The evolution of the colour bimodality is presented in Section 6 while in 
Section 7 we derive separate LFs for the populations of red and blue galaxies. 
Our conclusions are presented in Section 8.

Throughout this paper all magnitudes are given in the AB system 
(Oke \& Gunn 1983) and we have adopted a background cosmology 
with $\Omega_{\rm M}=0.3$, 
$\Omega_\Lambda=0.7$ and $H_0=70~{\rm km~s^{-1}}$.

\section{The Data-sets}
The data exploited in this paper cover a large range in wavelengths from the
optical to the mid-infrared and have been obtained
by combining three different surveys. The optical data were
taken with the Subaru Telescope as part of the Subaru/XMM-Newton Deep Survey
(Sekiguchi et al. 2005), the near-infrared data form part 
of the early-data release (Dye et
al. 2006) of the UKIRT Infrared Deep Sky Survey (Lawrence et al. 2006) and
mid-infrared data have been obtained with the {\it Spitzer} satellite as part of the 
{\it Spitzer} wide-area infrared extragalactic (SWIRE) survey (Lonsdale et al.
2003; 2004). 
In this Section the main properties of these three surveys are briefly outlined. 

\subsection{The Subaru/XMM-Newton Deep Survey}
The Subaru/XMM-Newton Deep Survey (SXDS) is a multi-wavelength survey covering
an area of $\sim 1.3$ square degrees. This area, centred on RA=02:18:00 and 
Dec=$-$05:00:00 (J2000), benefits from deep optical imaging undertaken with
Suprime-Cam  (Miyazaki et al. 2002) on Subaru. The 5 over-lapping Suprime-Cam 
pointings provide broad band photometry in the $BVRi'z'$ filters to typical 
$5 \sigma$ depths of $B=27.5$, $V=26.7$, $R=27.0$, $i'=26.8$ and $z'=25.9$ 
(within a $2$-arcsec diameter aperture).

The X-ray observations of this field were obtained  with the XMM-Newton 
satellite  and comprise 400 ksec exposures 
in 7 contiguous fields to a depth of 
$\rm \sim few \; 10^{-15} \; erg \; cm^{-2} \; s^{-1}$. 
The field also has deep 1.4 GHz radio observations from the VLA (Simpson et al.
2006a) and $\rm 850 \mu m$ sub-mm observations from SCUBA 
as part of the SHADES survey (Mortier et al. 2005; Coppin et al. 2006).

\subsection{The UKIDSS Ultra Deep Survey}\label{ukidss}
The central region of the SXDS is being observed at near-infrared wavelengths 
with the Wide-Field Camera (WFCAM - Casali et al. 2006,  in preparation) 
on UKIRT, as part of the UKIDSS Ultra Deep Survey (UDS). The UDS is the 
deepest of the five surveys that constitute the  UKIRT Infrared Deep Sky 
Survey (UKIDSS - Lawrence et al. 2006) and ultimately 
aims to cover 0.8 square degrees to a depth  $J_{AB} = 26$,
$H_{AB} = 25.4$ and $K_{AB} = 25$ over a seven year period. 
AB magnitudes in the UDS have been obtained by using calibrations from 
Hewett et al. (2006).
For the present study the first twelve hours of $J,K$ imaging, 
which constitute the early data release (EDR -- Dye et al. 2006) of the UDS, 
have been used. These data reach  
$5 \sigma$ limits of $J_{AB} = 22.5$  and $K_{AB} = 22.5$  within 
a $2$-arcsec diameter aperture. The details of the stacking 
and catalogue extraction are fully described in Foucaud et al. (2007).

\subsection{The SWIRE survey}

The {\it Spitzer} wide-area infrared extragalactic survey (SWIRE -- Lonsdale et al.
2003; 2004) is a large {\it Spitzer}
Legacy program which covers seven high-latitude fields with a total area 
of $\sim 65$ sq. deg. In this paper we exploit the {\it Spitzer} observations 
covering 
$9.1$ sq. deg. in the XMM-LSS field which cover the SXDS field. 
The source catalogue based on the data release 2 (Surace et al. 2005) includes 
sources with detections at IRAC $\rm 3.6 \mu m$ and  $\rm  4.5 \mu m$ 
wavelengths to a depth of $\rm \sim 10 \mu Jy$ $SNR \simeq 10$ in both bands 
( $\simeq 21.4$ in AB magnitudes).

\begin{figure*}
\center{{
\epsfig{figure=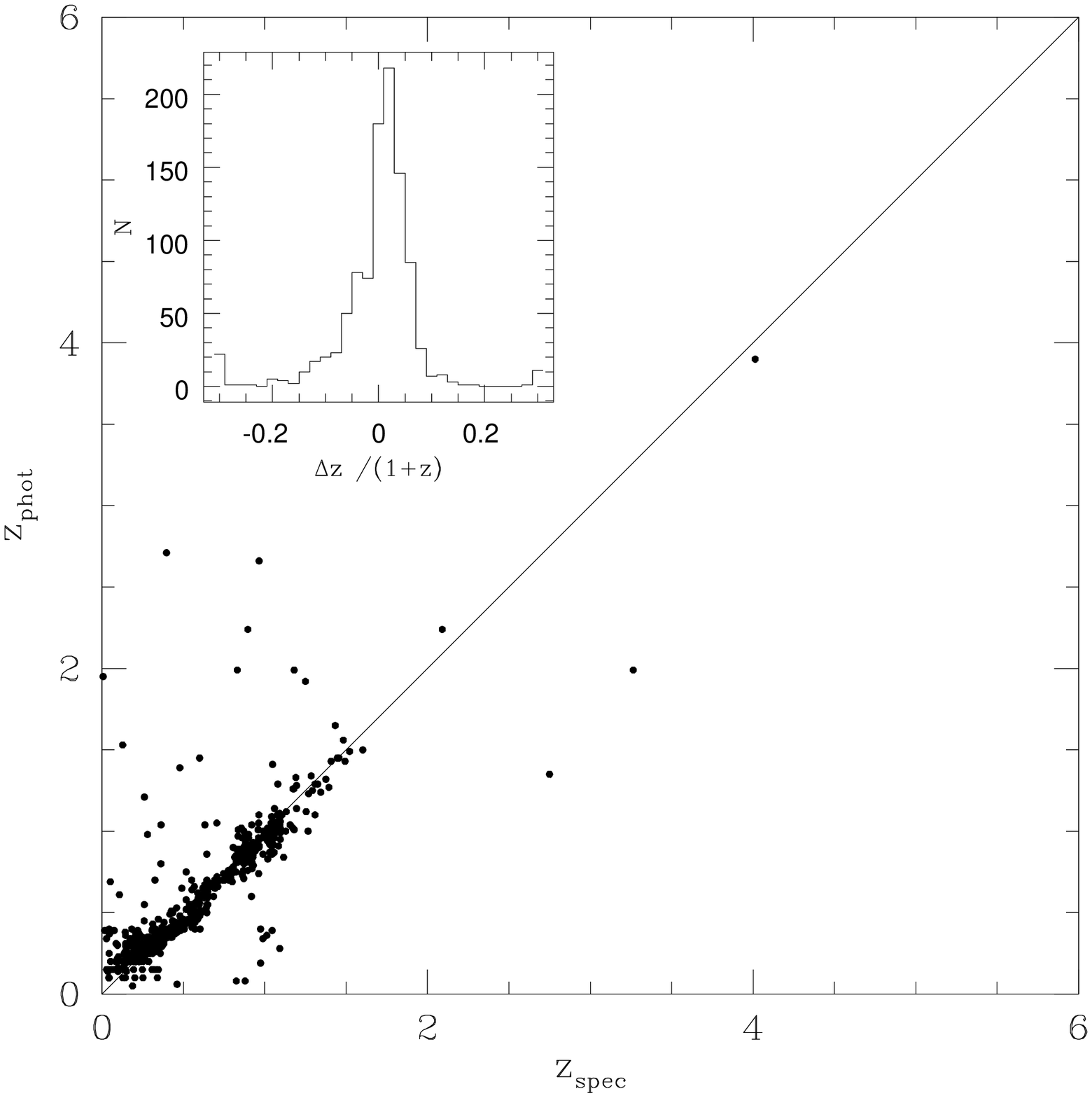, height=8cm}
\epsfig{figure=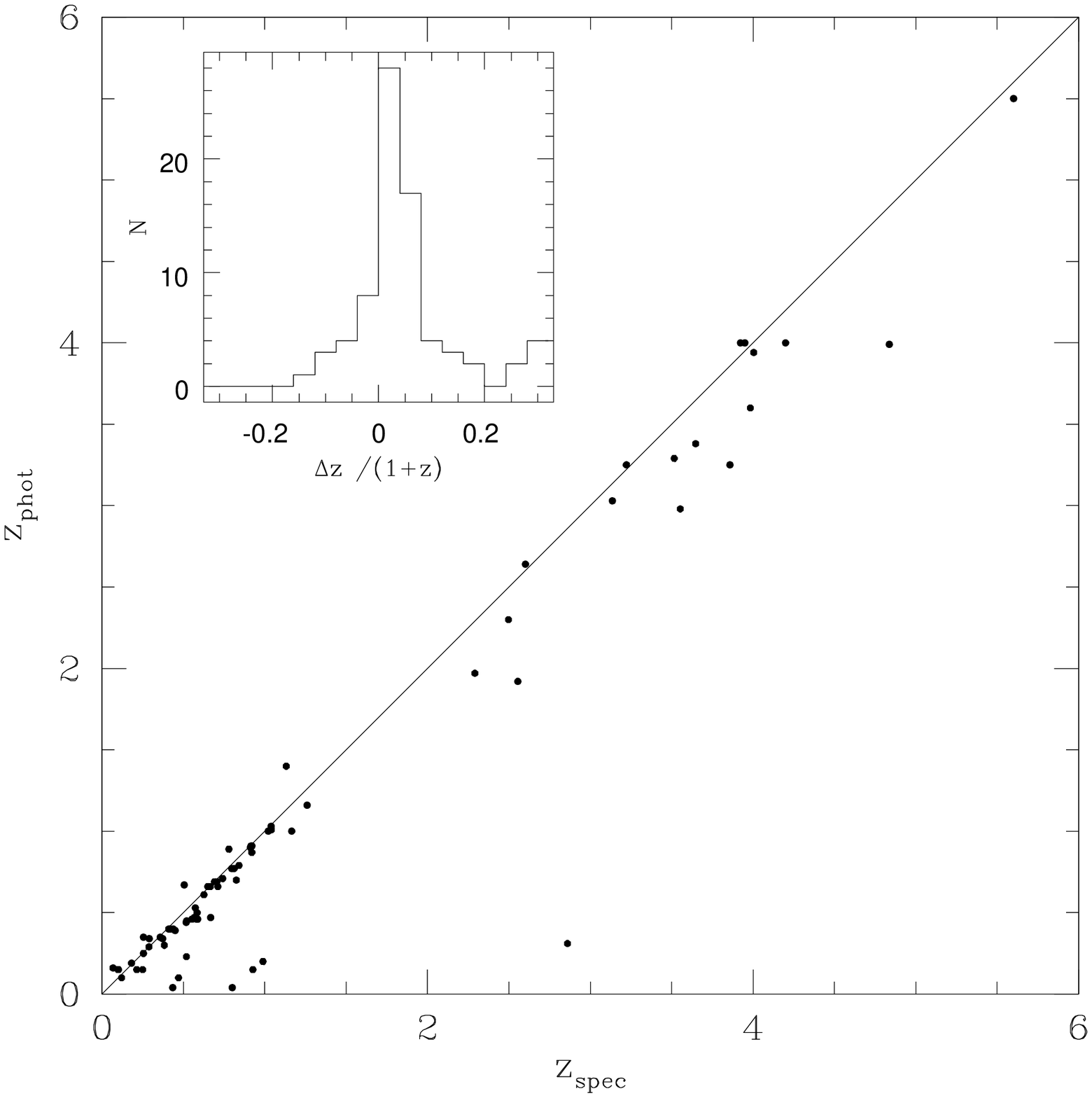, height=8cm}
\caption{\label{zz} Photometric redshift plotted versus 
spectroscopic redshift for $\sim 1100$
sources in the UDS sample with available spectroscopic redshift 
determinations (Yamada et al. 2005; Akiyama et al. in preparation; 
Smail et al. in preparation; Simpson et al. 2006a). Left panel shows the 
comparison for sources with $K_{AB} < 22$ while the comparison for 
fainter sources with $K_{AB} \ge 22$ is presented in the right panel. 
The small insets in both panels show the
distribution of the $\Delta z/(1+z)$.
The agreement is remarkably good with a mean value of 
$\Delta z/(1+z)$ of 0.006, and a standard deviation $\sigma =0.04$,  
excluding the clear outliers (see text for details).}
}}
\end{figure*}
\section{The UDS-galaxy sample}

Exploiting the multi-wavelength coverage and the large area provided by the
combination of the surveys described in the previous Section, we aim to derive
a complete sample of $K$-band selected galaxies, with photometrically
estimated redshifts. Only sources 
with \kab $\le 22.5$ have been included in the sample, which is the $5 \sigma$
magnitude limit of the EDR data (see Section \ref{ukidss}). 
The completeness at the limiting magnitude 
has been estimated to be above $70$\% with a fraction of spurious
detections less than 1\% (Foucaud et al. 2007). In order to have an
homogeneous multi-wavelength coverage for each object we use optical and
near-infrared photometry computed in a 2" diameter apertures PSF corrected to 
total. The consistent image quality of optical and near-infrared data ($0.7 <
\rm FWHM < 0.8$ arcsec) means that the differential aperture corrections between
bands are small ($\le 0.1$ mag). The Spitzer SWIRE photometry is in 2.8"
diameter apertures PSF corrected to total (see Surace et al. 2005).

Due to a small shift of the UDS field centre during the first year 
of observations, the current $J,K$ photometry in the EDR is uniform 
only over an area of $0.6$ square degrees. 
Therefore, in order to obtain homogeneous coverage over all 
the optical and near-IR spectral
range, we have limited our sample to include only sources in the 
central $0.6$ square degree area of the UDS field.

Many of cross-talk in the J and K-band images have
been masked in the process of image stacking and removed from the catalogue 
(see Foucaud et al. 2007). Moreover, given the surface density of optical 
sources and the 2" aperture diameters used
to compute the photometry we expect each cross talk to have  $<0.03$ counterparts
in the optical catalogues. All the sources with only J and K but not 
optical counterpart have been excluded from the final catalogue.

Excluding the saturated and masked regions (due to bright stars) our final 
sample comprises 25,564 sources with \kab $\le 22.5$.

All of the sources in this sample have optical and near-infrared photometry,
and a cross-match with the SWIRE catalogue also provided $\rm 3.6 \mu m$ and 
$\rm 4.5 \mu m$ fluxes for 14,879 sources (i.e. more than half of the sample).

\subsection{Photometric redshifts}

The photometric redshift for each galaxy in our UDS/SXDS sample has been 
computed by fitting the observed photometry ($ BVRi'z'JK$ and $\rm 3.6 \mu m$ 
and  $\rm 4.5 \mu m$ when available) with both synthetic and 
empirical galaxy templates. The fitting procedure is undertaken with a code
which is largely based on the public package {\sc HYPERZ} (Bolzonella, Miralles
\& Pell\'{o} 2000).
The stellar population synthesis models of Bruzual
\& Charlot (2003) were used to produce the synthetic templates, 
assuming a Salpeter initial mass function (IMF) with a lower and upper
mass cutoff of 0.1 and  100 $M_{\odot}$ respectively.  
We used a variety of star formation histories;
instantaneous burst and exponentially declining star formation with e-folding
times $\rm 0.3 \leq \tau (Gyr)\leq 15$, all with a fixed solar metallicity.
The dust reddening has been taken into account by 
following the obscuration law 
of Calzetti et al. (2000) within the range $0 \leq A_V \le 2$. We also added a
prescription for the Lyman series absorption due to the HI clouds in the inter
galactic medium, according to Madau (1995). Finally, at each redshift only
models with age less than the age of the Universe at that redshift have been
allowed.

As empirical templates we used the four Coleman, Wu and Weedman (CWW) 
observed spectra: Ellipticals, Sbc, Scd and Irr (Coleman, Wu \& Weedman 1980). 
In addition
we used three average templates obtained by the K20 survey 
(Cimatti et al. 2002);
red passive early-type galaxies, intermediate galaxies with emission lines but 
red continuum indices and  blue emission-line galaxies (Mignoli et al. 2005). 
An observed starburst SED from Kinney et al. (1996) was also included.
All of the empirical templates were extended into the ultraviolet and
infrared wavelengths by using the most appropriate synthetic model from Bruzual \& Charlot (2003).

Finally, a quasar template was also included to allow the photometric redshift 
code to identify  objects with type-I AGN features. 
For this purpose we used the composite
quasar spectra from the FIRST Bright Quasar Survey (Brotherton et al. 2001).

\subsection{Sources of contamination}

Since our primary goal is to construct a sample of $K$-band selected 
galaxies, it is important to clean our catalogue of stars and AGNs. 
To isolate stars we first used the Sextractor (Bertin \& Arnout 1996) 
stellarity parameter as measured in
the $i'$ band in the SXDS images. However, due to the $\simeq 0.8"$ FWHM
resolution of the available ground-based imaging $\sim 6500$ objects were
consistent with being unresolved point sources. 
Therefore, we combined the stellarity
information with the position of the stars in the BzK colour-colour diagram
(Daddi et al. 2004). As pointed out by Daddi et al., 
stars have colours that are
clearly separated from the region occupied by galaxies and can be efficiently
isolated with the criterion ${\rm (z' - K) < 0.3 (B - z') - 0.5}$. 
By applying this criterion, 
coupled with the requirement of stellarity parameter
$> 0.8$, we isolated $ 1990 $ stars (see Lane et al. 2007, for 
a full discussion of the features displayed by the BzK diagram for 
our new UDS/SXDS sample). It is worth noting that the vast majority
of these sources show an unacceptable value of the $\chi^2$ when fitted with 
the galaxy templates.

By using the quasar template in the SED fitting process we also isolated 
$909$ sources for which the photometry is dominated by AGN light and therefore 
better fitted by the AGN template. 
These objects are characterised by a blue UV-optical continuum which is  
virtually 
featureless in the broad-band photometry. Therefore, for these sources
the photometric redshift technique (which relies on the identification of 
continuum features, such as the $\rm 4000 \AA$ break in the case of galaxies)
is very hard to apply. 
The uncertainties related to the adopted quasar template from the FIRST survey 
(Brotherton et al. 2001) could in principle introduce some bias in the fitting
procedure, specially due to the contribution of galaxy light to the composite
template (see e.g. Glikman et al. 2006  and Maddox \& Hewett 2006 for a detailed
discussion). However, this is not a major issue for the present work since the
quasar template has only been used to isolate very blue featureless AGN
therefore excluded from the final galaxy sample. The uncertainties related to
the quasar template will however only affect $\simlt 1$\% of the total galaxy
sample. Moreover the surface density of such objects in the UDS/SXDS 
field ($\simeq 0.4 / \rm arcmin^2$) is in agreement with that obtained in the 
GOODS/CDFS. In fact, 
Caputi et al.  (2006), identified nearly 60 AGN in their $K$-selected sample down to 
\kab $\le 22.5$ over an area of
131 $\rm arcmin^2$. 

Finally we excluded from the final sample $624$ sources for which the SED 
fitting procedure provided a bad $\chi^2$. 
Specifically, given the number of data points and degrees of freedom, 
we reject all fits for which the value of the $\chi^2$ falls outside the 
99.7\% confidence interval of the corresponding $\chi^2$ distribution.
A visual inspection of the SED fits for these sources revealed that their
photometry is contaminated, mainly due to an erroneous cross-match between 
the optical and NIR catalogues or source blending.

\subsection{The final UDS-galaxy sample} 
In summary, the final sample of galaxies in the UDS/SXDS field (hereafter the
UDS-galaxy sample) consists of $22,041$ sources
with \kab $ \le 22.5$, over an area of 0.6 square degrees, excluding stars, 
AGN and sources with contaminated photometry. All the sources in the final
UDS-galaxy sample have optical + near-IR photometry and for nearly half of 
them we also possess $\rm 3.6 \mu m $ and $\rm 4.5 \mu m$ detections.

\begin{figure}
\center{{
\epsfig{figure=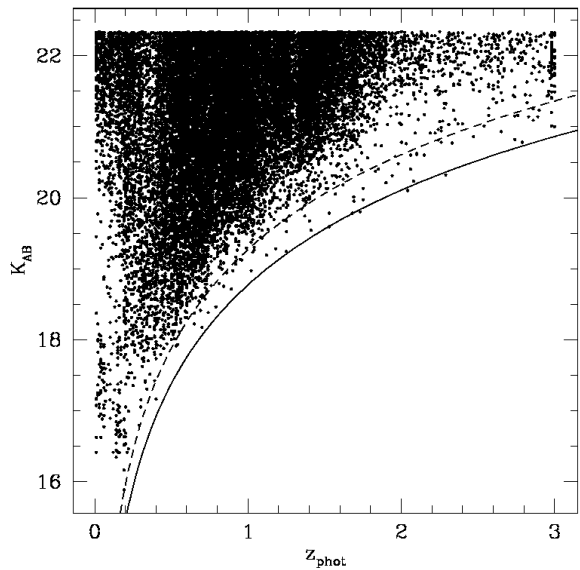, height=8cm}
\caption{\label{kz} A plot of observed $K$-band magnitude versus photometric redshift for the galaxies in the new UDS EDR sample.
The dashed line corresponds to the empirical $K - z$ 
relation for radio galaxies
as obtained by Willott et al. (2003), which approximately coincides with the
passive evolution of a $3L^*$ starburst formed at redshift $z_f =10$. The
solid line corresponds to the passive evolution of a $10L^*$ starburst.}
}}
\end{figure}

\section{The redshift distribution of the UDS-galaxy sample}
The SED fitting procedure provides two vital pieces of information
for each galaxy: the best fitting galaxy template and a photometric
redshift.
In this Section we check the reliability of
our photometric redshift estimates and compare the redshift distribution of
sources in the UDS-galaxy sample with that derived by previous studies in
different fields. 

\subsection{Spectroscopic redshifts}

Figure \ref{zz} shows the comparison of our photometric redshift estimates with
the spectroscopic values for $\sim 1100$ galaxies in the UDS field spanning 
$0 <z <6$ (Yamada et al. 2005; Akiyama et al. in preparation; Smail et al. in
preparation; Simpson et al. 2006a). We observe a good agreement between
estimated and real redshifts in the vast majority of the cases with only $< 3$\%
of clear outliers. The  $\Delta z/(1+z) \equiv  (z_{\rm spect} - z_{\rm phot}) / 
(1+z_{\rm spect}) $ has a mean of $ -0.003$ with a standard deviation 
$\sigma = 0.1$. By excluding the clear outliers (with $|\Delta z/(1+z)| >0.2$)
we obtain a mean of $ 0.006$  with a $\sigma$ reduced to $ 0.04$. This accuracy
is comparable to the best available from other surveys such as GOODS and COSMOS
(Caputi et al. 2006; Grazian et al. 2006; Mobasher et al. 2007). 
The residual small focusing of the
photo-z technique on the scale of $\delta z \simlt 0.05$ does not affect 
any of our results since our binning in redshift to compute LF and colour bimodality 
is at least of 0.25. 

The accuracy of our photometric redshift estimates is preserved also at faint
K-band magnitudes. In fact, $\simeq 10$\% of the sources with available spectra
have magnitude $K \ge 22$ and the distribution of the $\Delta z/(1+z)$ for these
sources has values for the mean and $\sigma$ consistent with the ones derived
for brighter sources (see right panel of Figure \ref{zz}).
 The statistics at those faint magnitudes is not large enough to
allow us to completely exclude systematics at $z > 2$ due to low signal-to-noise
in the $K$-band and specially in the Spitzer IRAC bands. However, it is worth
noting that for this work we focus at $z \simlt 2$, a range in which 
our photometric redshifts are more reliable, as shown in Figure \ref{zz}.

\begin{figure*}
\center{{
\epsfig{figure=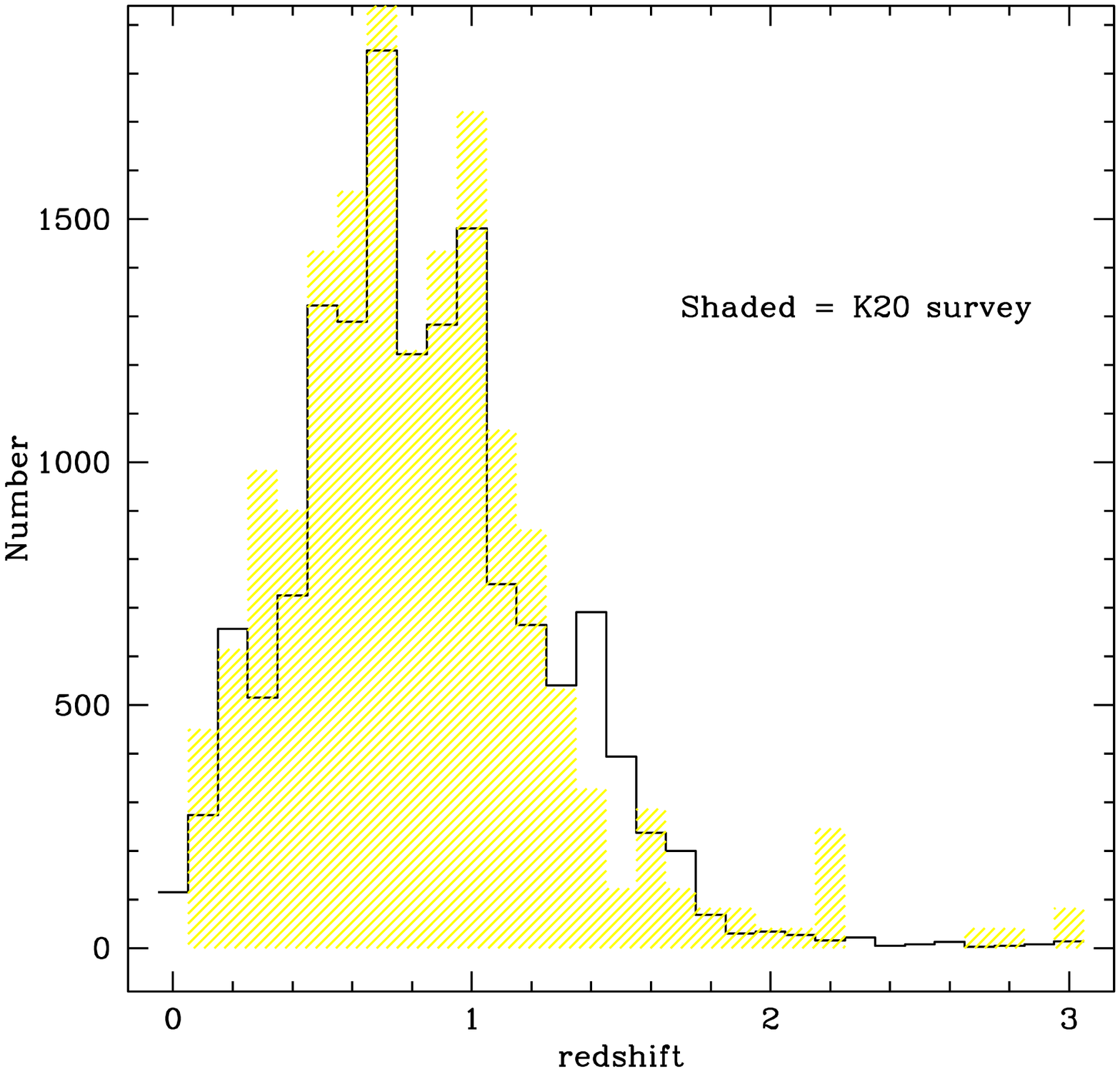, height=8cm}
\epsfig{figure=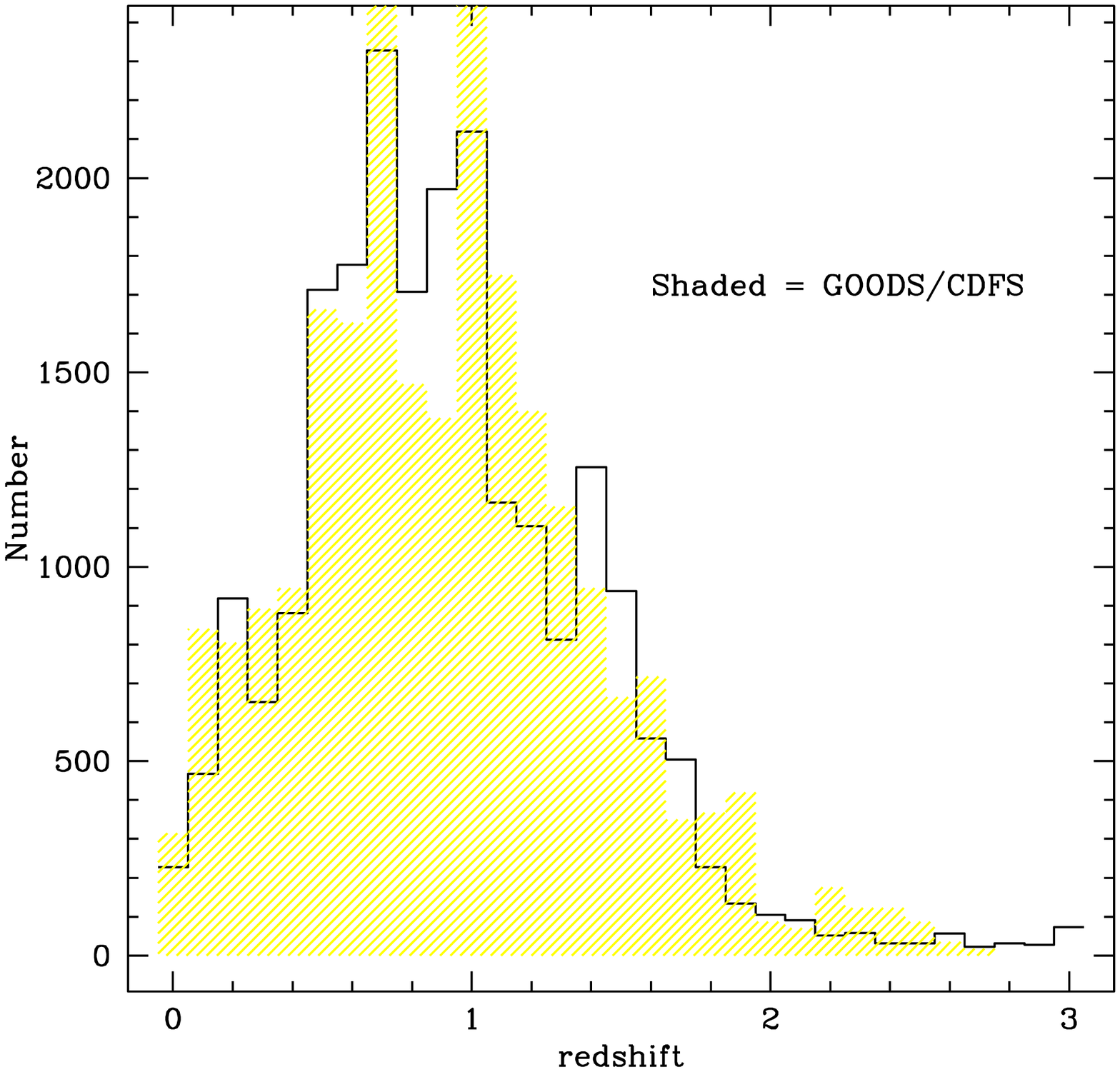, height=8cm}
\caption{\label{zdist} The estimated redshift distribution
of the new UDS galaxy sample, compared with previously published 
redshift distributions from near-infrared samples of comparable depth
but much smaller area. The solid histogram shows the 
distribution of photometric redshifts for sources in the UDS galaxy sample 
with \kab $\le 21.9$  and \kab $ \le 22.5$ in the left and right panels
respectively.  For comparison, the
redshift distributions of galaxies from the K20 survey (Cimatti et al. 2002;
Mignoli et al. 2005) and GOODS/CDFS (Caputi et al. 2006; Cirasuolo et al. in
preparation) are also  plotted as shaded
histograms in the left and right panels respectively, after 
normalisation to the same area as the UDS galaxy sample.}
}}
\end{figure*}

Finally, it is also worth noting the good accuracy of our photometric redshift
estimates at high redshift, $z \ge 4$. In particular, in the field there 
is a faint galaxy which has been detected in the $K-$band (\kab $=23.17$) 
with WFCAM and which has a very high spectroscopically  confirmed  
redshift of $z = 5.6$ (Ouchi et al. 2005). For this galaxy 
our analysis yields a photometric redshift of $z =  5.5 \pm 0.2$, 
providing additional confidence in the robustness of 
our redshift estimation procedure out to very high redshifts.

\subsection{The K-z relation}
Figure \ref{kz} shows the $K-z$ relation for the 
sources in the final UDS-galaxy sample. This plot illustrates the coverage 
of the $K-z$ plane provided by the sample, and also provides a useful 
sanity check that not only the overall redshift distribution, but also
the photometric redshift values for individual sources are
reasonable.

The dashed line represents the empirical $K-z$ relation for radio galaxies
as obtained by Willott et al. (2003), which approximately coincides with the
passive evolution of a $3L^*$ starburst formed at redshift $z_f =10$. The
solid line corresponds to the passive evolution of a $10L^*$ starburst with the
same formation redshift. To help minimise catastrophic errors in the 
photometric redshift estimation procedure, we introduced 
a prior which required that
the redshift of the sources should not exceed the $K-z$ relation of a passively
evolving $10L^*$ (i.e. the solid line). In fact this prior was only 
required for 12 galaxies, and it can be seen 
that the distribution of galaxies on the $K-z$ plane appears reasonable, 
with no significant excess of galaxies towards the bright 
limit.

\subsection{The BzK diagram}
Another consistency check on our photometric redshifts 
is offered by the BzK diagram.  
As expected, all the sources with photometric $z > 1.4$ lie  
in the regions with $BzK \equiv (z' - K) - (B-z') \ge -0.2 $  and 
$BzK < -0.2 \;
\cap \; (z' - K) > 2.5$ discussed by Daddi et al. (2004).
In fact due to sheer number of sources in the UDS EDR sample, the BzK 
diagram displays a number of interesting features of relevance to galaxy
evolution, some of which have not been evident in any previous studies.
The BzK diagram, along with a discussion of these features, is discussed
in detail by Lane et al. (2007).

\subsection{Comparison with previous studies}
The redshift distribution of our new UDS galaxy sample is shown in Figure
\ref{zdist}. The left-hand 
panel shows (as the solid histogram) 
the redshift distribution for the subset of sources with \kab  
$\le 21.9$ (which corresponds to \kv $ \le 20$) 
to facilitate comparison with the distribution 
obtained for the complete sample  
of 545 spectroscopically-confirmed sources (down to
\kv $= 20$) in the K20 survey 
(Mignoli et al. 2005). The right-hand panel instead
shows the redshift distribution of the UDS galaxy sample down to \kab $\le
22.5$ this time compared with the equivalent result obtained in the 
GOODS Chandra Deep Field South (CDFS) by Caputi et al. (2006) down to the same 
magnitude limit. 
Nearly half of the sources in the
GOODS/CDFS have spectroscopic redshifts and the remainder have photometric
redshifts computed by using 12 broad-band photometry from 
$\rm 0.4 \mu m$ to $\rm
8 \mu m$ (Cirasuolo et al. - in
preparation). Both the distributions from K20 survey and GOODS/CDFS have been
re-normalised to the same area as the UDS-galaxy sample. 

We find a very good agreement between the redshift distribution we 
obtain for sources in the UDS galaxy 
sample and the ones displayed by the smaller 
samples which, however, do obviously possess a much higher
fraction of 
spectroscopically confirmed redshifts.
It is also evident from Figure \ref{zdist} that the surface density of 
galaxies in the present sample is in agreement with those derived from 
these previous studies.
The consistency of the redshift distributions and of the surface densities 
at different magnitude limits provides further confidence 
in the reliability of our sample and our photometric redshift estimates.

\section{Luminosity Function}\label{lfsect}

\begin{figure*}
\center{{
\epsfig{figure=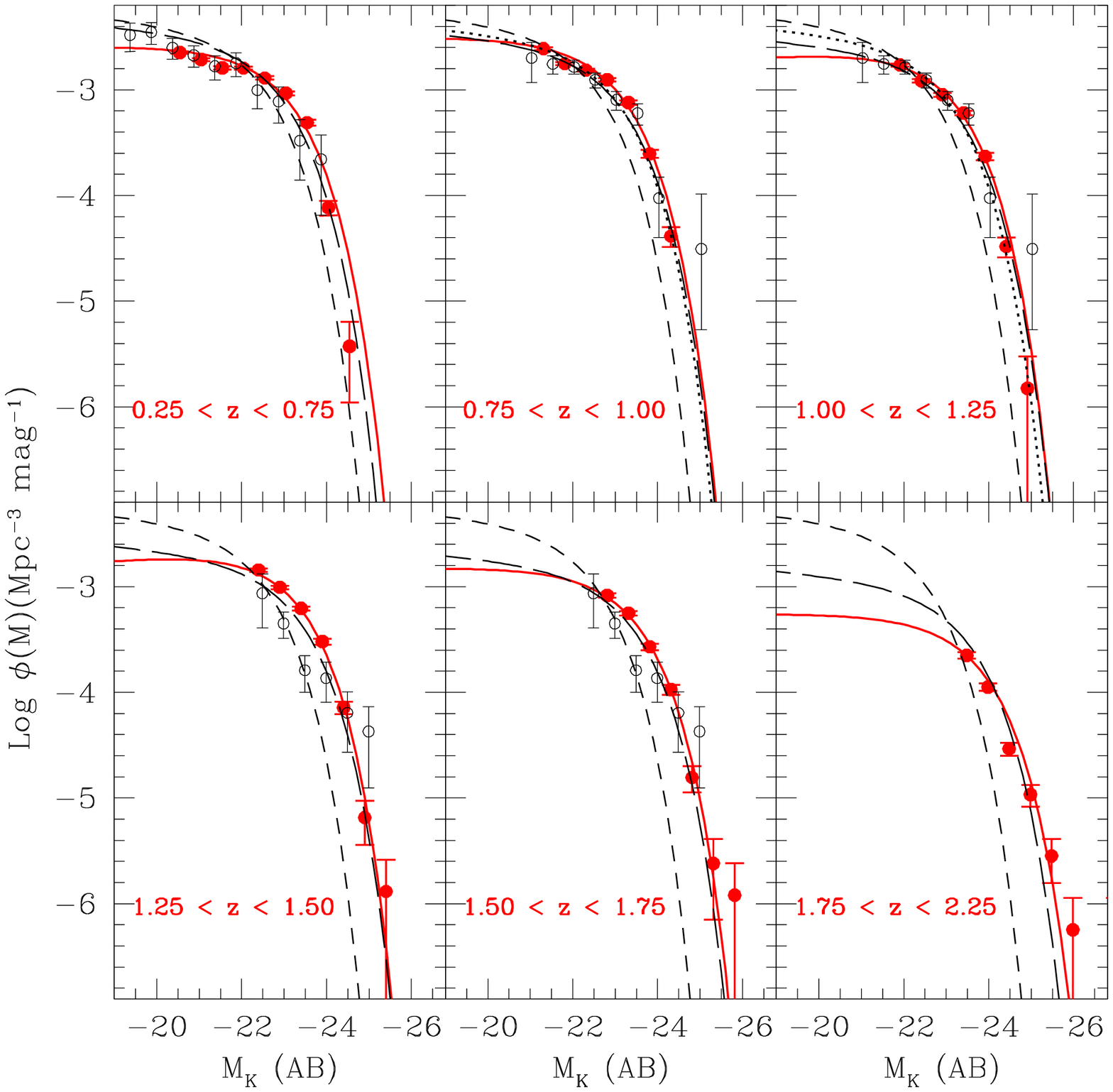, height=18cm}
\caption{\label{lf} Rest-frame $K$-band luminosity function in 
six redshift bins.
The solid dots show the LF obtained from the $\rm 1/V_{max}$ method for sources
in the UDS sample with \kab $ \le 22.5$. The open symbols correspond to the LF
obtained from the K20 survey (Pozzetti et al. 2003) in three redshift bins
([0.2-0.65], [0.75-1.3], [1.3-1.9]), 
while the solid line is the
best fit Schechter function obtained from the likelihood analysis. The LF
obtained from the GOODS/CDFS sample (Caputi et al.
2006) is indicated by the long dashed line. 
Also shown, for comparison, are the LFs at $z=0$ (dashed line) and 
$z=1$ (dotted
line) from Kochanek et al. (2001) and Drory et al. (2003), respectively.}
}}
\end{figure*}

In this Section we exploit the large area and wide redshift coverage 
of the UDS galaxy  sample to derive the evolution of the near-infrared
LF from the local Universe up to $z \simgt 2$.
The rest-frame $K$-band magnitudes for sources in the sample have been computed by
using the closest observed band 
depending on the redshift of the source. 
This method in fact minimises the uncertainties related to the  
$k$-corrections, even though in the $K$-band these corrections have a small variance
compared to the optical wavelengths (Cowie et al 1994; Poggianti 1997).
In our case the availability of the IRAC $3.6 \mu$m and $4.5 \mu$m
data enables us to directly trace the rest-frame $K$-band up to $z \sim 1$ 
for more than half of the sample. However, as a test we also compared the absolute $M_K$
magnitude obtained by using IRAC data with that obtained by directly
extrapolating the observed $K$-band magnitude. 
The distribution of the difference of the magnitudes computed with these 
two different methods is centred around zero with rms $\sigma \sim 0.35$.
Therefore, these two methods provide consistent results and for our dataset we
used the $M_K$ estimated by using the IRAC bands when available.
 
Figure \ref{lf} shows the rest-frame $K$-band luminosity function (LF) in the
redshift 
range $0.25 \le z \le 2.25$ computed in six redshift bins by using both the
classical $\rm 1/V_{max}$ method (Schmidt 1968) and the maximum likelihood
analysis (Marshall et al 1983). The $\rm 1/V_{max}$ method does not require any
assumption regarding  
the shape of the LF or parameter dependence but can be biased by
strong density inhomogeneity  in the field.  
The solid dots in Figure \ref{lf} represent the LF obtained with this method. 
Only the bins above the luminosity-completeness limit at each redshift bin are 
shown and the error bars on each point correspond to the Poissonian errors.

The solid line in Figure \ref{lf} shows the LF obtained in each redshift bin 
with the maximum likelihood method. This is a parametric method, for which we
chose a Schechter function to describe the shape of the LF:
\begin{equation}
\phi(M) = 0.4 {\rm ln(10)} \phi^* 10^{-0.4 \Delta M (\alpha +1)} \rm exp(-10^{-0.4\Delta M})
\end{equation} 

with $ \Delta M = M_K - M^*_K$. The overall normalisation ($\rm \phi^*$) has
been obtained by matching the number of observed galaxies in each redshift bin.
The best fit values for the parameters $\alpha$, $\rm M_K^*$ and $\phi^*$ in
each redshift bin are shown in Table \ref{tab_lf}. 

In the determination of the LF both in the $\rm 1/V_{max}$ method and maximum
likelihood analysis a correction factor has been applied to take into account
the incompleteness effects. In fact, as described in Foucaud et al. (2007), 
the UDS EDR sample is nearly 100\%  complete at \kab $\le 21.9$ and the 
completeness drops to $\sim 70$\% at \kab $= 22.5$.

Due to the limiting magnitude \kab $\le 22.5$ our ability to trace the faint 
end of the LF decreases with redshift. However, up to $z \sim 1.5$  the
likelihood analysis suggests that the value of the faint end slope 
$\alpha$ does not
change with redshift and stays consistent with the local value of 
$\alpha \sim 1$. The actual data do not allow us any estimate of the slope 
for $z \simgt 1.5$. 
Therefore, in the likelihood analysis at redshifts larger than $z=1.5$ we 
decided to fix the value
of the faint end slope to the mean value obtained at lower redshifts 
($\alpha = 1$). This is consistent with the results of deeper surveys 
(e.g. Caputi et al. 2006). 
Recently,  by exploiting a sample of $\sim 300$ objects 
selected from the Hubble deep field south (HDF-S) down to \kab $\le 24.9$ 
Saracco et al. (2006) found
the value of the slope of the LF to be constant with redshift and consistent
with the local value ($\alpha \sim 1$) up to $z \sim 3$. 

\begin{table}
\begin{center}
\caption{\label{tab_lf} Best fit parameters for the Schechter LF.}
\begin{tabular}{cccc} \hline \hline
z   &   $\alpha$  & $M^*_K$ & $\rm \phi^* (10^{-3} \; Mpc^{-3})$ \\ \hline \hline 
$0.25 - 0.75$  & $-0.99 \pm 0.04$     & $-22.86 \pm 0.08$  &  2.9 $\pm 0.2$ \\
$0.75 - 1.00$  & $-1.00 \pm 0.08$     & $-22.86 \pm 0.11$  &  3.4 $\pm 0.3$\\
$1.00 - 1.25$  & $-0.94 \pm 0.13$     & $-22.93 \pm 0.14$  &  2.8 $\pm 0.4$ \\
$1.25 - 1.50$  & $-0.92 \pm 0.18$     & $-23.03 \pm 0.16$  &  2.6 $\pm 0.3$ \\
$1.50 - 1.75$  & $\rm -1.00 \; fixed$ & $-23.23 \pm 0.09$  &  1.7 $\pm 0.2$ \\
$1.75 - 2.25$  & $\rm -1.00 \; fixed$ & $-23.58 \pm 0.13$  &  0.6 $\pm 0.3$ \\
\hline \hline
\end{tabular}
\end{center}
\end{table}

The agreement between the estimation of the LF obtained with the two independent
methods ($\rm 1/V_{max}$ and maximum likelihood) is evident in Figure \ref{lf}. 
We also found our LF to be consistent 
with the results derived by Pozzetti et al. (2003) from the K20 survey and Caputi 
et al. (2006) from GOODS/CDFS. It is worth noting that in this work we have
nearly  a factor of 10 more objects compared to the GOODS/CDFS sample and a factor
$\sim 40$ more than the K20 survey. Due to the large area we also have
enough statistics to be able to
properly trace the bright end of the LF over our full redshift range
with negligible cosmic variance.

The comparison with the local LF obtained by Kochanek et al (2001) from the Two
Micron All Sky Survey (2MASS) shown in Figure \ref{lf} confirms the substantial
brightening of the characteristic magnitude $M^*_K$ with redshift. In agreement
with previous studies (Pozzetti et al 2003; Drory et al. 2003; Feulner et al.
2003; Caputi et al. 2006; Saracco et al. 2006) we find $M^*_K$ to be
brighter by $\sim 1$ magnitude at $z \sim 2$ compared to the local Universe. 
The lack of data at the faint end of the LF does not allow us to precisely  
trace the density evolution of the LF. 
However, up to $z \sim 1.75$ where the likelihood estimate of the best fit
parameters for the LF is still reliable (see Table \ref{tab_lf}) we observe a
decrease in the space density by a factor $\sim 2$ compared to $z = 0$. This
decrement increases up to a factor $\sim 4.5$ if the highest redshift bin is
considered. This is in good agreement with the results obtained by  
Caputi et al. (2006) in the GOODS/CDFS as also shown in Figure \ref{lf}.
However, the much large statistics obtained with the current work allows for
substantial improvement in determining the LF, especially at its bright end.  
A more detailed likelihood analysis of the LF, including the 
parameterisation 
of the evolution and the determination of the faint end
slope up to the highest redshift bin will be possible with the next data 
release of the UDS which will reach \kab $\sim 23.5$.

\begin{figure*}
\center{{
\epsfig{figure=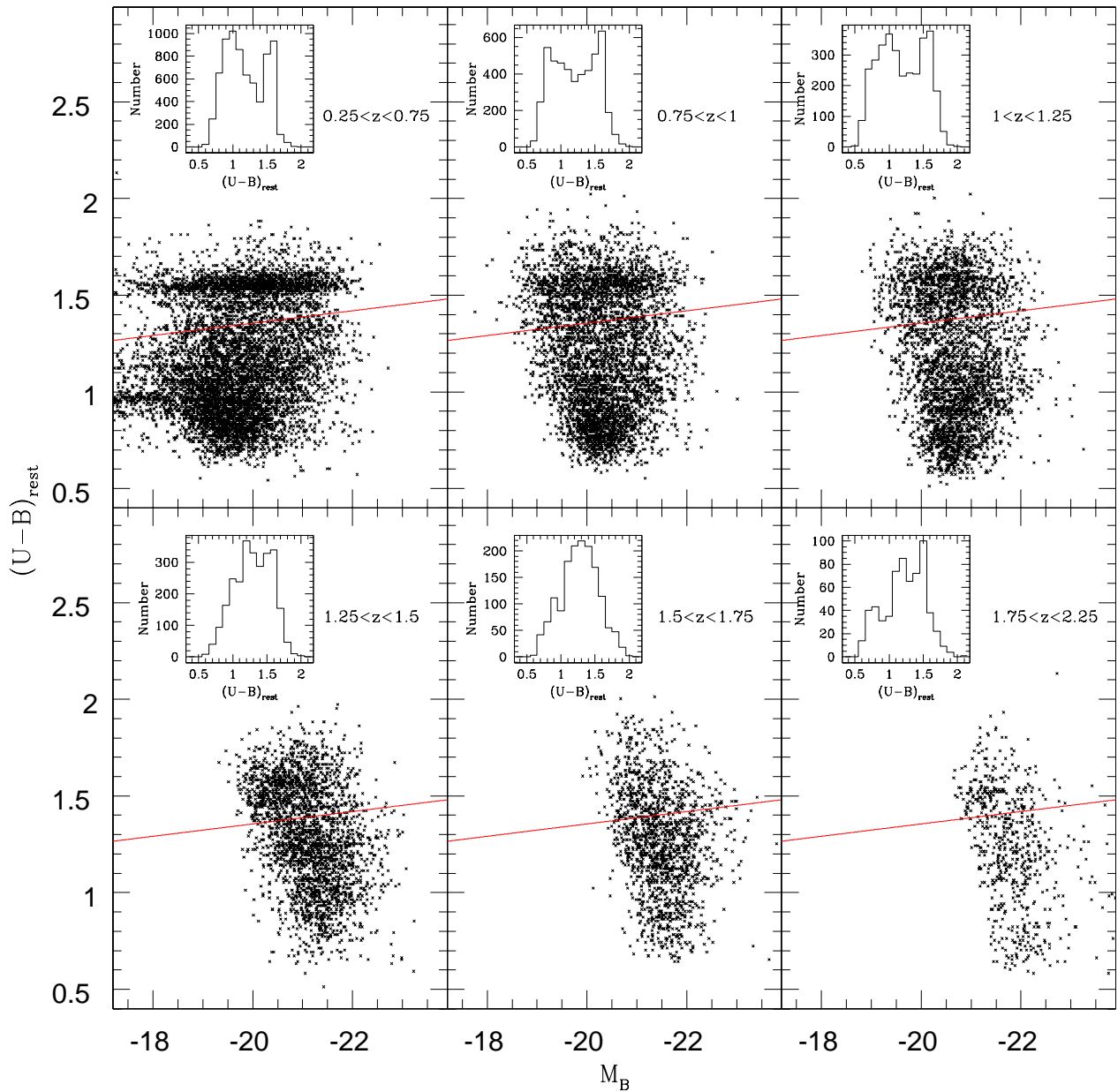, height=18cm}
\caption{\label{ub} Rest-frame $(U-B)$ colour versus absolute magnitude in the $B$
band in six redshift intervals for sources in the UDS galaxy sample. The small
insets show the distribution of the rest-frame ($U-B$) colour in each redshift
bin. The diagonal line represents the luminosity-dependent colour cut (see
Equation \ref{ccut}) adopted by
van Dokkum et al. (2000) to divide the populations of red and blue galaxies.}
}}
\end{figure*}

\section{The colour bimodality}
Following the above 
analysis of  the evolution with redshift of the population of 
$K$-band selected 
galaxies it is also interesting to investigate the role of the different 
galaxy 
populations and how they  contribute to the global evolution of the 
$K$-band LF. 
For this purpose we analysed the rest-frame $(U-B)$ colour of sources in the UDS-galaxy
sample; according to the well-studied local colour-magnitude relation
(e.g. Visvanathan \& Sandage 1977; Bower et al. 1992; Strateva et al. 2001) 
the red and blue galaxies show a bimodal distribution which allows one to
easily separate the population of red/early type galaxies from the blue/late type
systems. 

Figure \ref{ub} shows the rest-frame 
($U-B$) colour versus absolute magnitude in the $B$-band in the same six
redshift intervals used for the LF.  The rest-frame $U$ and $B$
band magnitudes have been computed by using the best fit template,
according to the redshift of the source.
The small insets in each
panel show the distribution of the rest-frame $(U-B)$ colour in each redshift bin.
In agreement with studies in the local Universe (Strateva et al. 2001; Hogg et al. 2002;
Blanton et al. 2003a,b; Baldry et al. 2004) and at $z \sim 1$ (Bell et al.
2004a; Weiner et al. 2005; Willmer et al. 2006; Franzetti et al. 2007) we find
that the colour-magnitude relation 
exhibits a clear bimodality up to $z \sim 1$.

The data also show the luminosity-dependent behaviour of the bimodality 
(Aragon-Salamanca et al. 1993; Stanford et al. 1995, 1998; Ellis et al. 1997;
Kodama et al. 1998; van Dokkum et al. 2000) 
with the brighter sources displaying redder
colours. In Figure \ref{ub}  the diagonal line represents the colour-magnitude 
relation derived by van Dokkum et al. (2000) for red galaxies in distant
clusters adding an offset of  -0.25 magnitudes to separate the red and blue 
populations (see Willmer et al. 2006):
\begin{equation}\label{ccut}
(U-B) = -0.032 (M_B + 21.52) + 1.4
\end{equation}  
converted to the AB magnitude system. As shown in Figure \ref{ub}, this
luminosity-dependent colour cut is in good agreement with the observed
bimodality present in the UDS galaxy sample, at least up to $z \sim 1.5$.

However, it is interesting to note that the strength of the colour bimodality fades with
increasing redshift and nearly disappears at $z \simgt 1.5$. 
The extend to which the apparent disappearance of the colour bimodality is
due to photometric errors will be revealed as the UDS data progressively
improve.
A more detailed study on the evolution of the bimodality
with redshift is beyond the scope of the present paper and will be
addressed in detail by Eales et al. (2007 - in preparation).

 However, it is worth noting 
the difficulty of the broad band rest-frame colours to distinguish between
the different populations of galaxies that show the same optical and
near-infrared colours, i.e.  
the old passive evolving sources and the dusty star-forming one.
As well demonstrated in a
recent paper by Stern et al. (2006) dusty starbursts and passively
evolving galaxies are virtually indistinguishable in the broad band photometry 
at wavelengths $\le 10 \mu$m.

In the local Universe, the population of red galaxies is mostly constituted by 
morphologically early type objects (Sandage \& Visvanathan 1978; 
Bower et al. 1992; Terlevich et al. 2001) and the mixing between the populations
of early and late type objects is relatively low. 
Recent studies on the
SDSS survey by Strateva et al. (2001) found only 
$\sim 20$\% of late type galaxies to have red colours compatible with the red
sequence. 

This low  contamination by the late type galaxies in the red sequence 
has also been observed at intermediate redshift ( $z < 1$) both in the field
(Kodama et al. 1999; Bell et al. 2004b) and in clusters (Couch et al. 1998; van
Dokkum et al. 2000; van Dokkum \& Franx 2001). 
These results are in good agreement with the well defined bimodality we observe 
at $z \simlt 1$ in the UDS-galaxy sample (see Figure \ref{ub}). 

However, the contamination and the mixing between early and late type galaxies
in the colour-magnitude diagram increases with redshift as clearly shown in
Figure \ref{ub}.
Recently, Franzetti et al. (2007) exploiting data at $z \sim 1$ 
from the VIMOS-VLT Deep Survey (VVDS) 
found that even though the red peak
of the colour distribution is mainly populated by early type galaxies and the
blue peak by late type ones, there is a strong contamination between the two
populations. In particular, they found that in the population of red galaxies
only $\sim 65$\% are early type galaxies, while the remaining $\sim 35$\% are late
type objects. 



\begin{figure*}
\center{{
\epsfig{figure=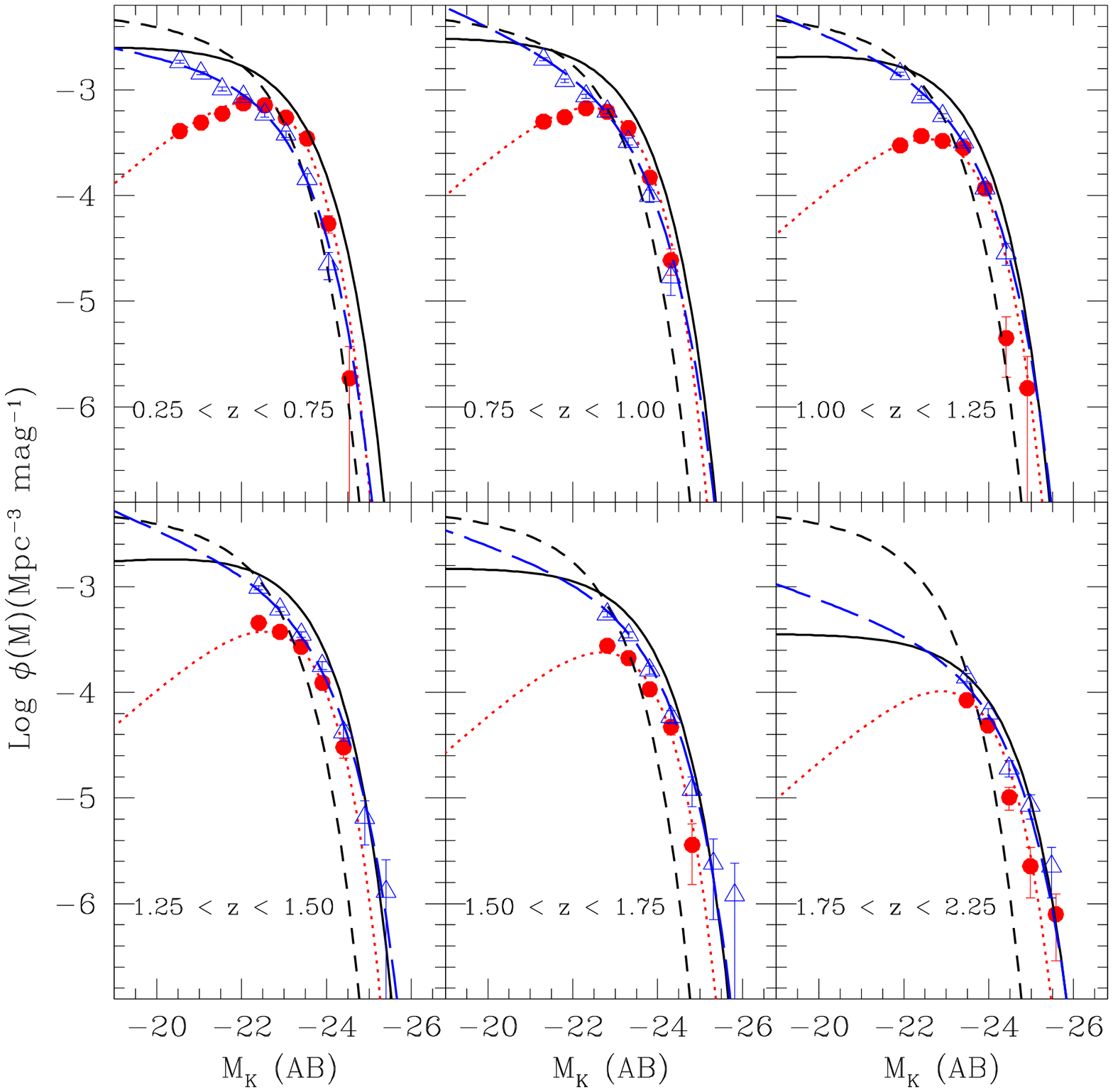, height=18cm}
\caption{\label{lf_rb} Rest-frame $K$-band LF for sources in the UDS-galaxy sample
sub-divided by colour according to Equation \ref{ccut}. The solid dots
represent the LF of the population of red galaxies while the open triangles 
show the LF of the blue objects. The dotted and long dashed lines show 
the LF obtained in each
redshift bin  by using the maximum likelihood analysis for the red and
blue population respectively (see Table \ref{tab_2}). 
The solid line is the LF of the whole $K$-band selected sample in each redshift
bin as plotted in Figure \ref{lf} and tabulated in Table \ref{tab_lf}. 
For reference we also indicate by a dashed line the local $K$-band LF from
Kochanek et al. (2001). }
}}
\end{figure*}

\section{The Luminosity Function subdivided by colour type}
As discussed in the previous Section and in Eales et al. (2007 - in prep.),
classification by colour into red and blue
sources provides an efficient means of 
identifying the population of early and late type
objects at $z \simlt 1$. However, as also outlined by Eales et al.  
the difficulties in successfully exploiting this
colour selection increase with increasing redshift as demonstrated by the
progressively disappearance of the colour bimodality at $z \simgt 1$.
Therefore, in the following Section we will study the evolution with redshift 
of the populations of red and blue galaxies bearing in mind that this
roughly corresponds to the evolution of the early and late type objects only 
at low and intermediate redshift. However, it is worth remembering that 
even at these redshifts the 
cross-contamination of 
each population can be relatively high ($\sim 20-30$\% ; 
Strateva et al. 2001; Ilbert et al. 2006; Franzetti et al. 2007).

\begin{table}
\begin{center}
\caption{\label{tab_2} Best fit parameters for the Schechter LF of red and blue
galaxies.}
\begin{tabular}{cccc} \hline \hline
z   &   $\alpha$  & $M^*_K$ & $\rm \phi^* (10^{-3} \; Mpc^{-3})$ \\ \hline \hline 
\multicolumn{4}{ c }{Blue objects} \\

$0.25 - 0.75$  & $-1.19 \pm 0.05$ & $-22.71 \pm 0.11$  &  $1.5 \pm 0.2$  \\
$0.75 - 1.00$  & $-1.44 \pm 0.15$ & $-23.05 \pm 0.18$  &  $1.3 \pm 0.2$  \\
$1.00 - 1.25$  & $-1.38 \pm 0.19$ & $-23.17 \pm 0.21$  &  $1.3 \pm 0.3$ \\
$1.25 - 1.50$  & $-1.43 \pm 0.24$ & $-23.40 \pm 0.25$  &  $1.0 \pm 0.4$ \\
$1.50 - 1.75$  & $\rm -1.35 \; fixed$ & $-23.45 \pm 0.15$  &  $0.9 \pm 0.3$ \\
$1.75 - 2.25$  & $\rm -1.35 \; fixed$ & $-23.75 \pm 0.20$  &  $0.25 \pm 0.3$ \\ \hline
\multicolumn{4}{ c }{Red objects} \\

$0.25 - 0.75$  & $-0.12 \pm 0.1$ & $-22.36 \pm 0.07$  &  $2.3 \pm 0.2$  \\
$0.75 - 1.00$  & $-0.09 \pm 0.2$ & $-23.48 \pm 0.13$  &  $2.0 \pm 0.3$ \\
$1.00 - 1.25$  & $-0.05 \pm 0.4$ & $-22.65 \pm 0.17$  &  $1.0 \pm 0.3$ \\
$1.25 - 1.50$  & $-0.14 \pm 0.5$ & $-22.65 \pm 0.24$  &  $1.0 \pm 0.4$\\
$1.50 - 1.75$  & $\rm -0.1 \; fixed $ & $-22.81 \pm 0.10$  &  $0.7 \pm 0.3$ \\
$1.75 - 2.25$  & $\rm -0.1 \; fixed $ & $-23.04 \pm 0.21$  &  $0.2 \pm 0.4$ \\
\hline \hline
\end{tabular}
\end{center}
\end{table}

As a first step we sub-divided the UDS 
galaxy sample adopting the luminosity-dependent colour   
cut shown by Equation \ref{ccut} (van Dokkum et al. 2000). The resulting LFs 
for
the red and blue galaxies are shown in Figure \ref{lf_rb}. As for the LF of 
the total sample of $K$-band selected galaxies (see Section \ref{lfsect})
we used  both the $\rm 1/V_{max}$ and maximum 
likelihood methods. The best fit parameters for
the likelihood are shown in Table \ref{tab_2}. 
Due to the limiting magnitude \kab $\le 22.5$ the determination of the best
fit parameters in the redshift bin $1.75 \le z \le 2.25$ is particularly
challenging. Therefore, in the following discussion we consider only 
the redshift bins in the range $0.25 \le z \le 1.75$ and simply comment on the 
behaviour in the highest z bin. 

The limiting magnitude \kab $\le 22.5$ also does not allow us to obtain a 
precise
determination of the faint end of the LF, as already discussed in Section
\ref{lfsect}. Therefore, the maximum likelihood analysis has been 
performed both
leaving  $\alpha$ as a free parameter as well as fixing this value to the one 
obtained in the lower redshift bins, where our data better trace the faint-end
slope. The mean value for the parameter $\alpha$ in the redshift range $0.25
\simlt z \simlt 1.5$ is -1.35 and -0.1 for the blue and red objects 
respectively. Therefore, at $z \ge 1.5 $ we fixed  the slope of the LF at these
values. As is clearly seen in Figure \ref{lf_rb}, 
the population of blue galaxies exhibits a much steeper slope compared
to the red galaxies up to $z \sim 1.5$ in agreement with previous works (Lilly
et al. 1995; Blanton et al. 2001; Pozzetti et al. 2003;
Giallongo et al. 2005; Ilbert 2006).

\begin{figure}
\center{{
\epsfig{figure=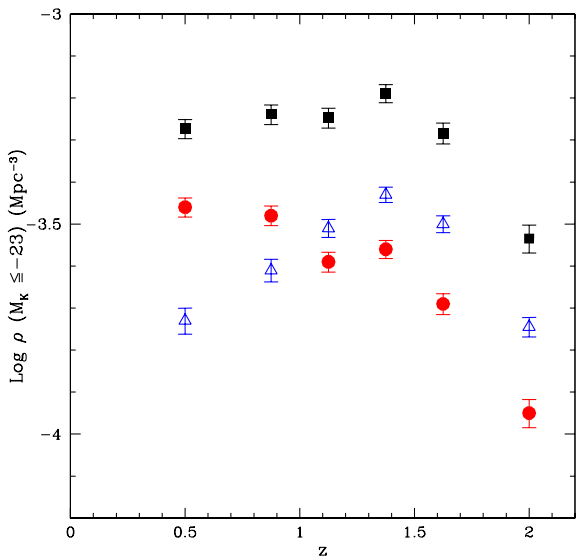, height=8cm}
\caption{\label{sd} Space density as a function of redshift for bright sources
with  $M_K \le -23$
classified as red (solid dots) and blue (open triangles) according to their 
rest-frame colours (see Equation \ref{ccut}). The black solid squares represent the
space density of all sources with $M_K \le -23$.}
}}
\end{figure}

The redshift
evolution of the blue and red populations also appears to differ.
Even though both the red and blue sources experience a brightening 
of the characteristic magnitude $M^*_K$, the blue objects show a slightly 
stronger evolution with $\Delta M^*_K \sim -0.7$ over the redshift range $0.25
\le z \le 1.75$ compared with $\Delta M^*_K \sim -0.5$ of the red population.
According to the results of the maximum likelihood analysis, 
both populations show a small
density evolution at $z \simlt 1$ while at higher redshifts the decrease in
number density is stronger for red objects compared with the blue ones. 
However, it is worth remembering that the uncertainties in the classification 
of the two populations based purely on the rest-frame colour cut expressed by
Equation \ref{ccut} can introduce severe bias, specially at higher redshift 
where the bimodality fades away. 

The inter-connection between the luminosity and density evolution coupled with
the different faint-end slope of the LFs for the two populations  
result in the predominance of the blue objects at faint magnitudes ($M_K \simgt
-22$) over the whole redshift range explored in this work. 
On the other hand, the bright end of the LF is dominated by red objects 
(see  Figure \ref{lf_rb}) but only up to $z \sim 1$. 
In fact, as shown in Figure \ref{sd} the space density of bright sources with 
$M_K \le -23$  is nearly constant for red objects in the range $0.25 \simlt z
\simlt 1.5$, while it increases by a factor $\sim 2$ for the blue ones over the
same redshift interval. 
The result is that the population of the brightest objects mostly consists  
of red sources at low and intermediate redshift, while at
$z \simgt 1$  there is an inversion in the proportion with a slight
predominance of blue galaxies.

It is also worth noticing that the mild density evolution of the red galaxies 
between $z=0$ and $z \simeq 1$ is in
good agreement with the recent results by Yamada et al. (2005). These authors found
that the number density of old bright passive galaxies is decreased by only 
a small factor (less than 2) compared with the local Universe.
This is  inconsistent with the results of Bell et al. (2004a) which instead 
suggested a decrease in the population of red galaxies by nearly a factor 10 
over the same redshift range. 
However, as shown by Bundy et al. (2006) this decrement can be due to a
selection  effect. In fact, these authors pointed out how the $R$-band limit 
--  the same band used by Bell et al. to select their sample -- 
introduces a bias against red galaxies, especially at $z \simgt 1$, and 
which introduces an artificially large 
decrease with redshift in the red population.  

\begin{figure}
\center{{
\epsfig{figure=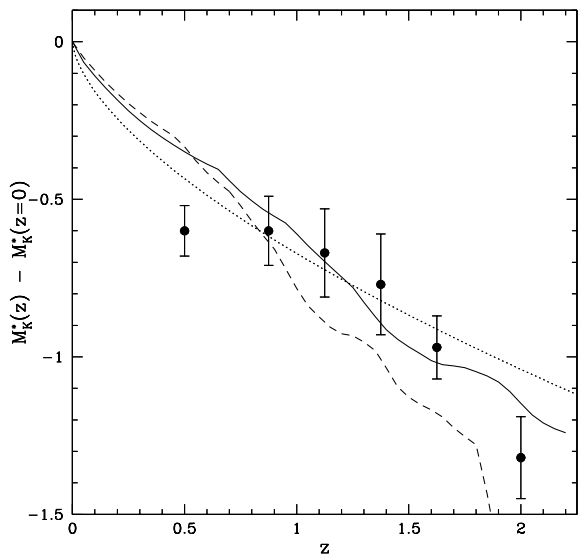, height=8cm}
\caption{\label{evol} The solid dots represent the evolution of the characteristic 
luminosity $M_K^*$ with redshift derived from the likelihood analysis relative to the 
one in the local Universe ($M_K^* (z=0) =22.26$)
as obtained by Kochanek et al. (2001). The dotted line is the evolution of $M_K^*$
derived by Caputi et al. (2006) from GOOD/CDFS. The solid and dashed lines represent
the evolution with redshift of the $K$-band absolute magnitude 
derived by a single burst model from Bruzual \& Charlot (2003) with formation
redshifts $z=3$ and $z=6$, respectively.}
}}
\end{figure}

\section{Discussion and Conclusions}
We have presented new results on the 
evolving near-infrared galaxy luminosity function, based on 
the analysis of a large sample of $K$-band selected galaxies
drawn from the Early Data Release of the UKIDSS Ultra Deep Survey. 
The complete 
sample includes $\sim 22,000$ sources down to \kab $\le 22.5$ over an area of 0.6
square degrees. Exploiting the complementary optical and mid-IR data provided by the
Subaru/XMM-Newton Deep Survey and {\it Spitzer}-SWIRE Survey we have 
obtained reliable
photometric redshift estimates for the sources in the sample. This allowed us to 
study the evolution of the $K$-band LF in the redshift range  
$0.25 \simlt z \simlt 2.25$. In agreement with
previous studies, we find that 
the characteristic luminosity of the near-infrared luminosity
function 
exhibits a substantial brightening of $\simeq 1$ magnitude between the local Universe
and $z \simeq 2$, while the total density decreases by a factor $\simeq 2$ over the
same redshift interval.

We have also used the rest frame $(U-B)$ colour to separate the sample into red and blue
galaxies. We find the 
classical bimodal colour distribution to be clearly present
at $z \simlt 1$. However, the strength of the colour bimodality is
found to decrease with increasing redshift and virtually disappears by 
$z \simgt 1.5$. The progressively improvement of the UDS data will assess
how the errors on the photometry affect this result.

By using the colour - magnitude relation to split the sample we have
analysed the
cosmological evolution of the red and blue galaxy populations. The blue 
objects clearly dominate the faint end of the LF over the whole redshift range
explored in this work. On the other hand, up to $z \simeq 1$ 
the brightest sources are mainly provided by red objects. 
In fact, the space density of red sources with $M_K \le -23$ is approximately
constant with redshift while the blue galaxies, with the same bright absolute
magnitude, exhibit a strong increase in their space density with redshift. 
Therefore, this different evolutionary behaviour of the two populations 
determines 
the preponderance of blue objects in the bright-end of the luminosity function at
redshifts $z \simgt 1$.

Compared with previous studies, 
the unique combination of large area and 
depth of the UDS minimises the effect of the cosmic variance and has led to a
substantial improvement in the determination of the LF. 
In particular, the large
area coverage has allowed us to accurately trace the evolution of the brightest
sources at $z \lsim 2$ with a high level of statistical
significance. As shown in Figure \ref{sd}, the space density of the total population
of bright sources ($M_K \le -23$) is roughly constant over the redshift range 
$0.5 \simlt z \le 1.5$. This strongly suggests the vast majority of the present-day
bright elliptical galaxies were already in place at $z \simeq 1.5$.

On the other hand, our analysis has also shown a change in the fraction of red and
blue galaxies with redshift. The majority of bright sources at $z \simlt 1$ exhibit red
colours consistent with having an old stellar population formed at high redshift. 
However, the increase of the fraction of blue galaxies with increasing redshift
-- indicative of an increase of the ongoing star formation -- suggests the major
formation epoch of the these bright sources to be at $z \simgt 1.5$.

However, it is worth noting that the colour classification based on the rest-frame ($U-B$) 
colour is not very accurate for distinguishing between
the different populations of galaxies that show the same optical colours, i.e.  
the old passive evolving sources and the dusty star-forming one.
In fact, studies of  the population of
Extremely Red Objects (EROs) at redshifts $1 \simlt z \simlt 2$ 
revealed this population to be
composed by both old passive evolving galaxies and dusty star-bursts, roughly in
the same proportion (Cimatti et al. 2002; Smail et al. 2002; 
Yan \& Thompson 2003; Cimatti et al. 2003; Gilbank et al., 2003; 
Moustakas et al. 2004; Simpson et al. 2006b). 

On the other hand, not all the blue objects are experiencing their major event of star
formation. In fact, it
only requires  less than 10\% of the total mass of an early type galaxy to
be involved in a recent episode of star formation to completely alter the    
colour of the objects and show it as a blue galaxy (Zepf 1997). 
This is because even a 
relatively small number of bright young stars dominates the emission at 
optical wavelengths. Moreover, a recent study from the VVDS in the redshift range $ 0.4
\simlt z \simlt 0.8$ by Ilbert et al (2006) found that $\sim 30$\% of bulge
dominated sources exhibit blue colours, indistinguishable from the late type
population (see also Im et al. 2001; Menanteau et al. 2004; Cross et al. 2004).

It is also interesting to notice that the brightening of the characteristic luminosity
($M_K^*$) of the whole $K$-band selected LF is consistent with the passive evolution of
a single stellar population. 
Figure \ref{evol} shows the evolution of 
$M_K^*$ derived from our likelihood analysis relative to $M_K^*$ at z=0
(Kochanek et al. 2001). Within the errors, our data agree remarkably well with 
the evolution of the $K$-band absolute
magnitude displayed by a single burst model from Bruzual \& Charlot (2003) with formation
redshifts in the range $3 \simlt z \simlt 6$. This result, coupled with the constant space
density of bright sources up to $z \simeq 1.5$, suggests the major assembly epoch of
bright sources to be at $z \simgt 2$, even though spread over a large range in
redshift. In fact, as shown from Figure \ref{sd}, nearly 40\% of the bright sources at
$z \simeq 1.5$ exhibit red colours. Assuming that half of them are dusty starbursts
-- as suggested by studies of ERO populations (e.g. Cimatti et al. 2002) -- 
we end up inferring that at least 20\% of sources are old passively evolving galaxies at these
redshift. Therefore,
this implies that a reasonable fraction of bright sources at $z \simeq 1.5$ was
assembled at very high redshift with very efficient star formation.

\section{acknowledgements}
MC, SF and CS would like to acknowledge funding from
PPARC.
RJM, OA and IS would like to acknowledge the funding of the Royal
Society.  We are grateful to the staff at UKIRT and Subaru for making these 
observations possible. We also acknowledge the Cambridge Astronomical
Survey Unit and the Wide Field Astronomy Unit in Edinburgh for processing the 
UKIDSS data.

\end{document}